\documentclass[%
 aip,
 amsmath,amssymb,
 reprint,%
]{revtex4-1}

\usepackage{graphicx}
\usepackage{dcolumn}
\usepackage{bm}

\usepackage[utf8]{inputenc}
\usepackage[T1]{fontenc}
\usepackage{mathptmx}
\usepackage{float}

\begin{document}

\preprint{AIP/123-QED}

\title[Synchronization and spatial patterns in forced swarmalators]{Synchronization and spatial patterns in forced swarmalators}

\author{Joao U. F. Lizarraga}
\affiliation{Department of Electrical and Mechatronics Engineering, Universidad de Ingenier\'ia y Tecnolog\'ia, Lima 15063, Lima, Peru \looseness=-1}
\author{Marcus A. M. de Aguiar}%
 \email{aguiar@ifi.unicamp.br}
\affiliation{Instituto de F\'isica 'Gleb Wataghin', Universidade Estadual de Campinas, Unicamp 13083-970, Campinas, SP, Brazil \looseness=-1}

\date{\today}

\begin{abstract}
Swarlamators are particles capable of synchronize and swarm. Here we study the effects produced by an external periodic stimulus over a system of swarmalators that move in two dimensions. When the particles are fixed and interact with equal strength (Kuramoto oscillators) their phases tend to synchronize and lock to the external stimulus if its intensity is sufficiently large. Here we show that in a system of swarmalators the force also shifts the phases and angular velocities leading to synchronization with the external frequency. However, the correlation between phase and spatial location decreases with the intensity of the force, going to zero in what appears to be a second order phase transition. In the regime of zero correlation the particles form a static symmetric circular distribution, following a simple model of aggregation. Interestingly, for intermediate values of the force intensity, a different pattern emerges, with the particles splitting in two clusters centered at opposite sides of the stimulus' location, breaking the radial symmetry. The two-cluster pattern is stable and active, with the clusters slowly rotating around the source while exchanging particles. 

\end{abstract}

\maketitle

\begin{quotation}
The behavior of many systems of coupled oscillators, such as neurons and fireflies, can be described by the Kuramoto model. The model illustrates how collective synchronization emerges as the intensity of the coupling increases past a critical value. A classic example is the synchronous flash of fireflies that arises as the night falls. Spatial collective organization is also possible and occurs frequently in flocks of birds or schools of fish. This type of swarming behavior occurs as a response to individual  repulsion to close neighbors and attraction to the group. The swarmalators model combines the main features of synchronization and swarming, defining a set of active particles that interact according to their phase values and spatial positions. Recently, the Kuramoto model was extended to describe systems that are driven by an external periodic stimulus, such as circadian rhythms or the flashing of artificial  lights. In this work we consider the effects of a periodic stimulus acting on swarmalators and study its consequences to the swarming patterns and phase dynamics. 
\end{quotation}

\section{\label{sec:introduction}Introduction}

The study of synchronization dates back to the famous observation of C. Huygens in the seventeenth century that two pendulum clocks hanging from a wall would swing in phase opposition after a short transient time \cite{oliveira2014huygens}. Much later, in 1967, Winfree \cite{winfree1967biological} introduced the first mathematical model based on coupled oscillators to study circadian rhythms. The system was then simplified and solved by Kuramoto in 1975 in the limit of infinitely many oscillators \cite{kuramoto1975international}. One of the most important features of Kuramoto's model was the identification of a second order phase transition to synchronisation as the intensity of the coupling between the oscillators increased. In the past 20 years the original Kuramoto model was extended and generalised to describe a number of interesting natural and artificial phenomena, such as fireflies \cite{moiseff2010firefly}, metronomes \cite{boda2013kuramoto}, neuronal systems \cite{mackay1997synchronized, gray1994synchronous}, electrochemical oscillators \cite{kiss2008}, explosive (first order) synchronisation \cite{Rodrigues2016} and externally forced systems of oscillators \cite{ott2008low, childs2008stability, moreira2019global, moreira2019modular, climaco2019}.

More recently the Kuramoto model was combined with models of swarming \cite{o2017oscillators}, that describe certain types of animal movement observed in flocks of birds, fireflies and schools of fish \cite{couzin2007collective, herbert2016understanding}. In this way, the effect of one oscillator over the other depends not on a previously established network of interactions, but on the distance between them, that might change dynamically. Phase synchrony, in turn, affects the intensity of the spatial coupling, changing how the oscillators align their velocities, for example. Such particles were termed swarmalators and one of the models proposed in \cite{o2017oscillators} is described by the equations

\begin{align} \label{eq:non-forcedx}
\dot{\vec{x}}_{i}&= \frac{1}{N} \sum_{j\neq i}^{N} \bigg[\frac{\vec{x}_{j} - \vec{x}_{i}}{|{\vec{x}_{j} - \vec{x}_{i}}|} (A+ Jcos(\theta_{j}- \theta_{i}))- B\frac{\vec{x}_{j} - \vec{x}_{i}}{|{\vec{x}_{j} - \vec{x}_{i}}|^2} \bigg] \\ \label{eq:non-forcedtheta}
\dot{\theta}_{i}&= \frac{K}{N} \sum_{j\neq i}^{N} \frac{sin(\theta_{j}- \theta_{i})}{|\vec{x}_{j}- \vec{x}_{i}|}. 
\end{align}

\noindent The phases $\theta_i$ represent the internal degree of freedom of the oscillators and correspond to the variables of the Kuramoto model. Here, contrary to the usual Kuramoto system, the intensity $K$ of the phase coupling is modulated by the spatial distance between the oscillators. The spatial motion is described by the particles velocities, which tend to align at short distances (with intensity $A$) and repel at large distances (parameter $B$). Attraction, furthermore, is amplified by phase coherence (parameter $J$). For $A$ and $B$ fixed the values of
$J$ and $K$ divide the behavior of the system into five different phases according to the spatial arrangement and phase distribution of the particles\cite{o2017oscillators}, as illustrated in the left panel of Fig. \ref{fig:F0F2}.

In this paper we are interested in the response of swarmalators to external perturbations. This is motivated by recent studies of periodically forced Kuramoto model where the competition between spontaneous synchronization (without the external force) and collective oscillation driven by the force (when its amplitude is sufficiently large) results in a rich dynamical structure with different scenarios separated by bifurcation curves\cite{ott2008low,childs2008stability,moreira2019global}. The periodic external force might represent, for example, day-night cycles, or an external stimulus to the neural system, that needs to be processed by the synchronous firing of neurons in the brain\cite{moreira2019modular}. Here we consider the effects of an external source of periodic perturbations over the phase of the swarmalators, such as a pulsing light placed in the middle of a cloud of fireflies. We show that the force tends to increase the phase synchronization of the particles, shifting the phase velocities towards the external frequency, but decreases the correlation between phase and spatial location, as illustrated in the right panel of Fig. \ref{fig:F0F2}. The transition between states fully synchronized with the external force and with no correlation between phase and spatial location to partially correlated states appears to be a second order phase transition. We also show that, depending on the model parameters and force intensity, a state where the particles divide into two clusters form. This new state, not found in the non-forced system, is stable and active, with the clusters rotating slowly around the origin and continuously exchanging particles.

This paper is organized as follows: in section II we introduce the forced swarmalators model and in section III we define the order parameters used to describe the system phases. We then review the main results of the free swarmlator system and describe the effect of the external force in terms of its amplitude and period. A brief discussion is then presented in section V.

\section{The forced swarmalators model}

In order to study the response of swarmalators to stimuli we consider the addition of a periodic external force to the original model described in Eqs. (\ref{eq:non-forcedx}) and (\ref{eq:non-forcedtheta}). The forced model is described by

\begin{align} \label{eq:dX}
\dot{\vec{x}}_{i}&= \frac{1}{N} \sum_{j\neq i}^{N} \bigg[\frac{\vec{x}_{j} - \vec{x}_{i}}{|{\vec{x}_{j} - \vec{x}_{i}}|} (1+ Jcos(\theta_{j}- \theta_{i}))- \frac{\vec{x}_{j} - \vec{x}_{i}}{|{\vec{x}_{j} - \vec{x}_{i}}|^2} \bigg] \\ \label{eq:dTheta}
\dot{\theta}_{i}&= F\frac{cos(\Omega t - \theta_{i})}{|{\vec{x}_{0} - \vec{x}_{i}}|} + \frac{K}{N} \sum_{j\neq i}^{N} \frac{sin(\theta_{j}- \theta_{i})}{|\vec{x}_{j}- \vec{x}_{i}|}  , 
\end{align}
where $F$, $\Omega$ and $\vec{x}_{0}= (x_{0}, y_{0})^T$, are amplitude, frequency and spatial location of the stimulus, respectively. We have fixed $A=B=1$ so that variations in the behavior of the forced system governed by Eqs. (\ref{eq:dX}) and (\ref{eq:dTheta}) depend only on parameters $J$, $K$, $F$, and $\Omega$.

The external stimulus is introduced in such a way to have direct effects only in the phases of the particles, not on their spatial motion. Therefore, although the stimulus originates from  a fixed position in space, there are no attractive or repulsive forces related to it: the particles can pass through the position of the stimulus with no explicit effects in their spatial dynamics. There are, however, indirect effects over the spatial dynamics mediated by the coupling term proportional to $J$, as shown by Eqs. (\ref{eq:non-forcedx}) and (\ref{eq:dX}). Also, as the effect of the stimulus on the phase dynamics is directly proportional to the amplitude $F$, and inversely proportional to the distance between the stimulus and the particle, the closer the particles are to the position of the stimulus the stronger the effects and vice versa.  

Eqs. (\ref{eq:dX}) and (\ref{eq:dTheta}) are invariant under the transformations $\theta_i \rightarrow \theta_i +\pi$ and $F \rightarrow -F$, which implies that the sign of $F$ should not matter.

\section{Order Parameters}
To measure the level of synchrony in the system, we define the usual complex order parameter
\begin{equation}\label{op: op01}
Z= Re^{\mathrm{i}\Psi}=\frac{1}{N} \sum_{j= 1}^{N}e^{\mathrm{i}\theta_{j}},
\end{equation}
It represents the center of mass of  particles phases distributed in the unit circle. Thus, if the system is fully synchronized ($\theta_{i} \cong \theta_0$), the norm $R$ tends to $1$, otherwise $R$ tends to $0$.

The level of correlation between angular position $\phi_{i}=tan^{-1}(y_{i}/x_{i})$, and phase $\theta_{i}$, can be quantified using another complex order parameter\cite{o2017oscillators}
\begin{equation}\label{op: op02}
W_{\pm}= S_{\pm}e^{\mathrm{i} \Theta_{\pm}}=\frac{1}{N} \sum_{j= 1}^{N}e^{\mathrm{i}(\phi_{j} \pm \theta_{j})}.
\end{equation}
When the norm $S= max(S_{+}, S_{-})$ is $1$ there is full correlation and, as the correlation decreases, $S$ tends to $0$. 

\section{Numerical Analysis}

\subsection{Free swarmalators}

\begin{figure*}
\includegraphics[scale= 0.3]{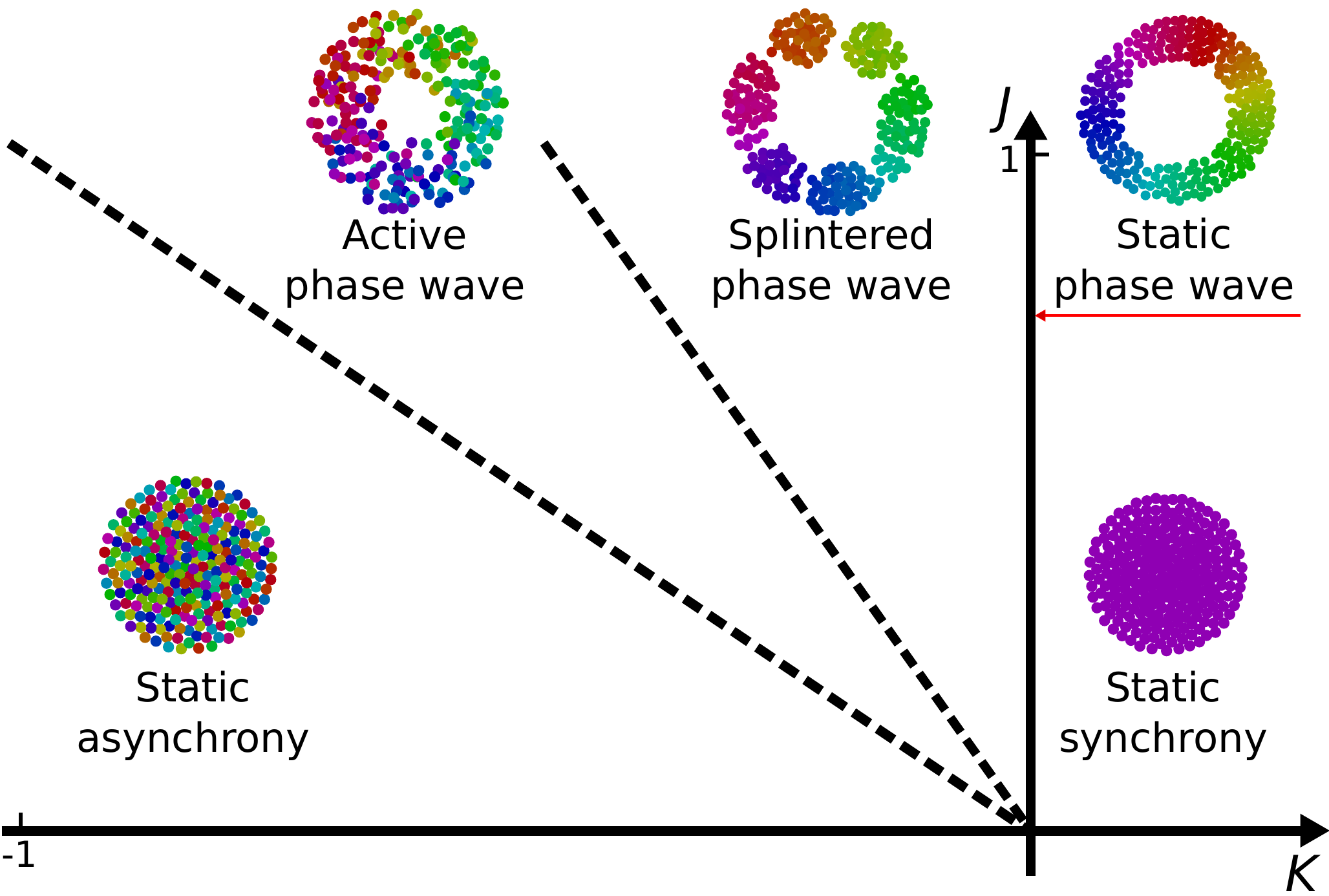} \qquad \qquad
\includegraphics[scale= 0.3]{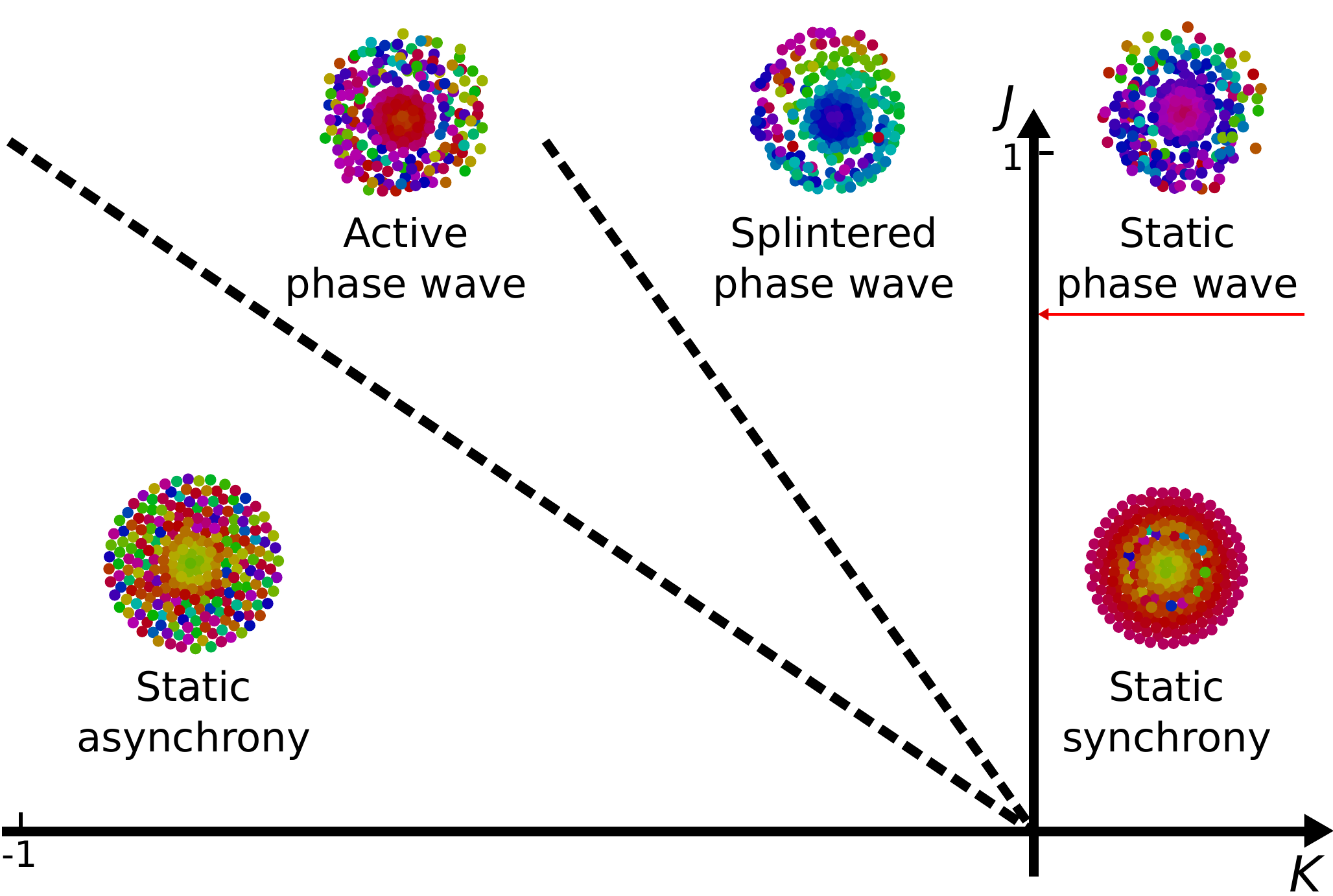}
\caption{\label{fig:F0F2} (left) Schematic phase diagram of free swarmalators showing the five phases of the system and a snapshot for each case.(right) Phase diagram and snapshots for $F=2$ and $\Omega = 3\pi/2$. In both cases the colors represent the phase of the oscillators.}
\end{figure*}

Before we describe the results of the forced system, we briefly review the behavior of free swarmalators\cite{o2017oscillators}. As illustrated in the left panel of Fig. \ref{fig:F0F2} the free system displays five different phases depending on the values of $J$ and $K$. Table 1 lists the five phases, together with the values of the parameters $J$ and $K$ used to construct the snapshots shown in the figure. These were obtained by integrating the equations of motion, starting with random initial conditions, until equilibrium. All numerical simulations were performed  using the MATLAB function `ode45', and unless otherwise stated, the number of particles is $N =250$. Colors represent the phase of the particles. Two static phases display circular spatial distributions, whereas the other phases display a ring-like structure with stronger correlation between location and phases. The only `active' phases (where particles keep moving at long times) are the active phase wave and splintered phase wave.

\begin{figure*}
\includegraphics[scale= 0.4]{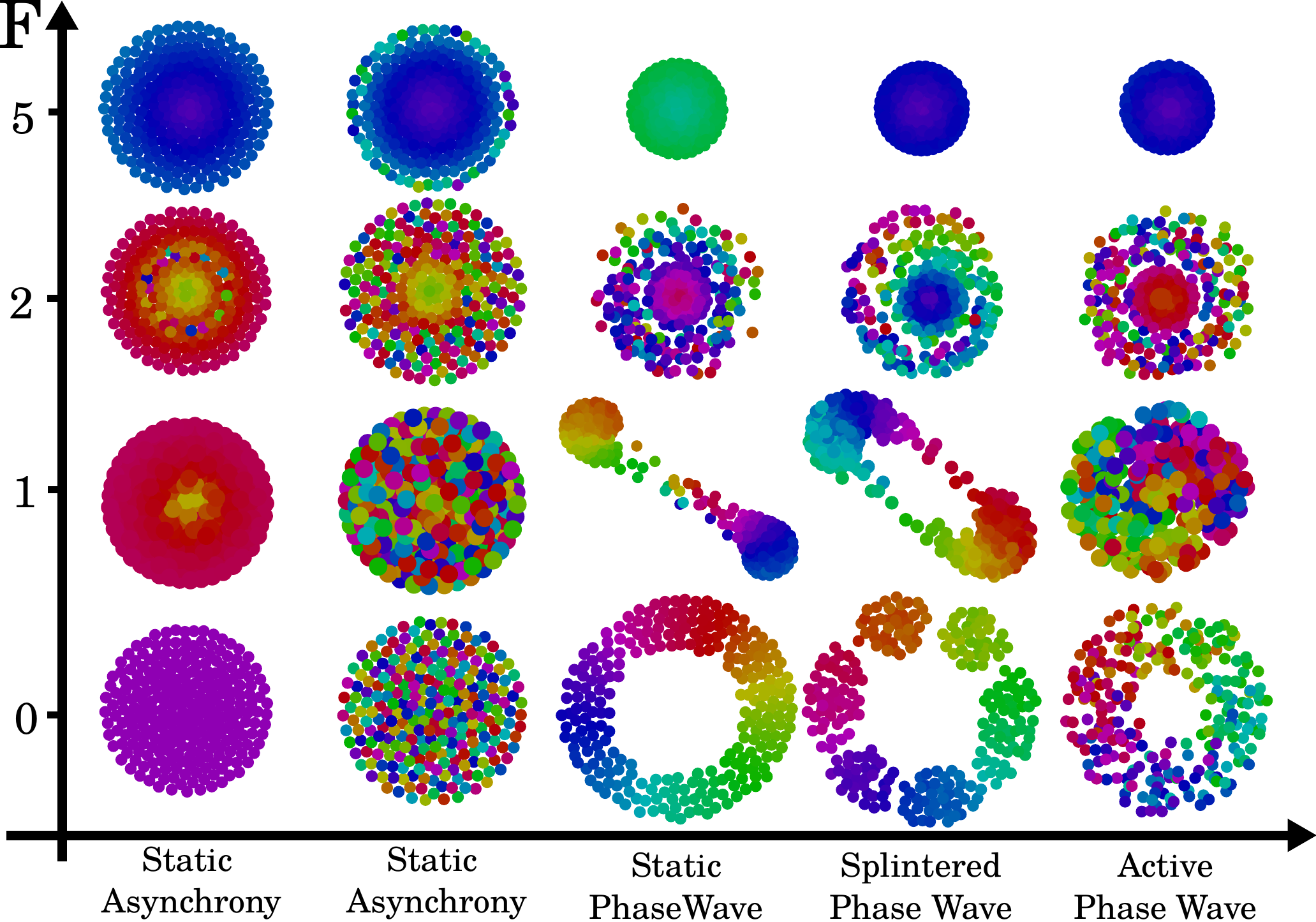}
\caption{\label{fig:transition} Equilibrium states for each of the five phases as a function of the external force amplitude $F$ and $\Omega = 3\pi/2$. The parameters $J$ and $K$ for each phase are listed in  1. }
\end{figure*}

\subsection{Forced swarmalators}

Simulations were performed in order to analyse transitions and long-term states in the forced swarmalators system. For each simulation the parameters $(J, K)$ were chosen as in Table 1. A qualitative view of role of $F$ is shown in the right panel of Fig. \ref{fig:F0F2} for $F=2$. Although the force acts only on the oscillator's phases, it indirectly attracts the oscillators towards its center. This happens because, as the force amplitude increases, the oscillators tend to synchronize with the external frequency $\Omega$, increasing the contribution of the attractive term proportional to $J$ (see below). 

Fig. \ref{fig:transition} shows how each of the five phases responds to $F$. The spatial distribution of the particles in the two circular static phases shrink only slightly, but the ring-like structures contract significantly, as the value of $J$ in these cases are larger. The most striking spatial pattern occurs for the SPW and SPpW and $F=1$, where the cloud of particles splits in two clusters at opposite sides of the force source. This is a completely different type of state, as it breaks the radial symmetry. The clusters slowly rotate around the origin while exchanging particles. When the force is further increased the clusters merge again in the center and the radial symmetry is recovered. 

Movies showing the dynamical behavior of the particles as a function of the external force for all five phases are available in the supplemental material (SM). In the case of the SA phase with the source placed at the center of the initial cloud of particles (movie 1), the particles close to the source quickly synchronize with the stimulus forming a small circular cluster that flashes together. As the force increases the size of the cluster also increases until it encompasses the entire system. 

The dynamics for the SPW and SPpW are more complex, since there are no particles initially close to the source (movies 2 and 3). In these cases for $F=0.5$ the particles close to the source start to move around it clockwise, whereas those at the outer boundary of the ring move counter clockwise. As the force intensity increases to $F=1$ the particles divide in two clusters opposing each other. The clusters rotate slowly around the source while exchanging particles. For $F > 1$ the two clusters are attracted back to the source eventually synchronizing and forming a small circle around it, similar to what is seen in the SA transition, but displaying complex oscillatory dynamics midway to the final state. The two-clusters state is stable for $F=1$, and not a transient, as can be seen in movie 4. 

When the location of the external stimulus is placed off center, the dynamics of synchronization follows a corresponding asymmetric behavior, as
the source affects more strongly the particles in its neighborhood first. This is illustrated in movie 5 for the SPW state, which shows the formation of the two opposing clusters of particles similar to what is seen in movie 2. 

The transitions in the APW are similar to the SPW, but without the splitting in two groups (movie 6). A simulation with only 5 particles is also shown in movie 7. Finally, a second example of dynamics with the source placed at the periphery of the initial cloud of particles is shown in movie 8 for the SA phase. In this case the interplay between phase sync and spatial behavior is seen more clearly: particles close to the source synchronize with the force and cause other particles away from it to oscillate, without breaking the initial circular symmetry of the system. As the force increases more particles sync with the group surrounding the source, until the whole system is included.

\begin{table}[]
\caption{Phases displayed by swarmalators, values of parameters used in numerical simulations and qualitative descriptors of transitions in \textbf{S} and \textbf{R} when the system is forced (\textit{s} for smoother transitions and \textit{u} for sharper transitions).}
\begin{tabular}{llcccc}
\hline
\multicolumn{2}{c}{\textbf{State}}    & \textbf{J} & \textbf{K} & \textbf{S} & \textbf{R} \\ \hline
\textit{SS}   & Static Synchrony      & 0.1        & 1          & \textit{s} & \textit{u} \\
\textit{SA}   & Static Asynchrony     & 0.1        & -1         & \textit{s} & \textit{u} \\
\textit{SPW}  & Static Phase Wave     & 1          & 0          & \textit{u} & \textit{u} \\
\textit{SPpW} & Splintered Phase Wave & 1          & -0.1       & \textit{u} & \textit{u} \\
\textit{APW}  & Active Phase Wave     & 1          & -0.75      & \textit{u} & \textit{u} \\ \hline
\end{tabular}
\end{table}

\begin{figure*}
\includegraphics[scale= 0.385]{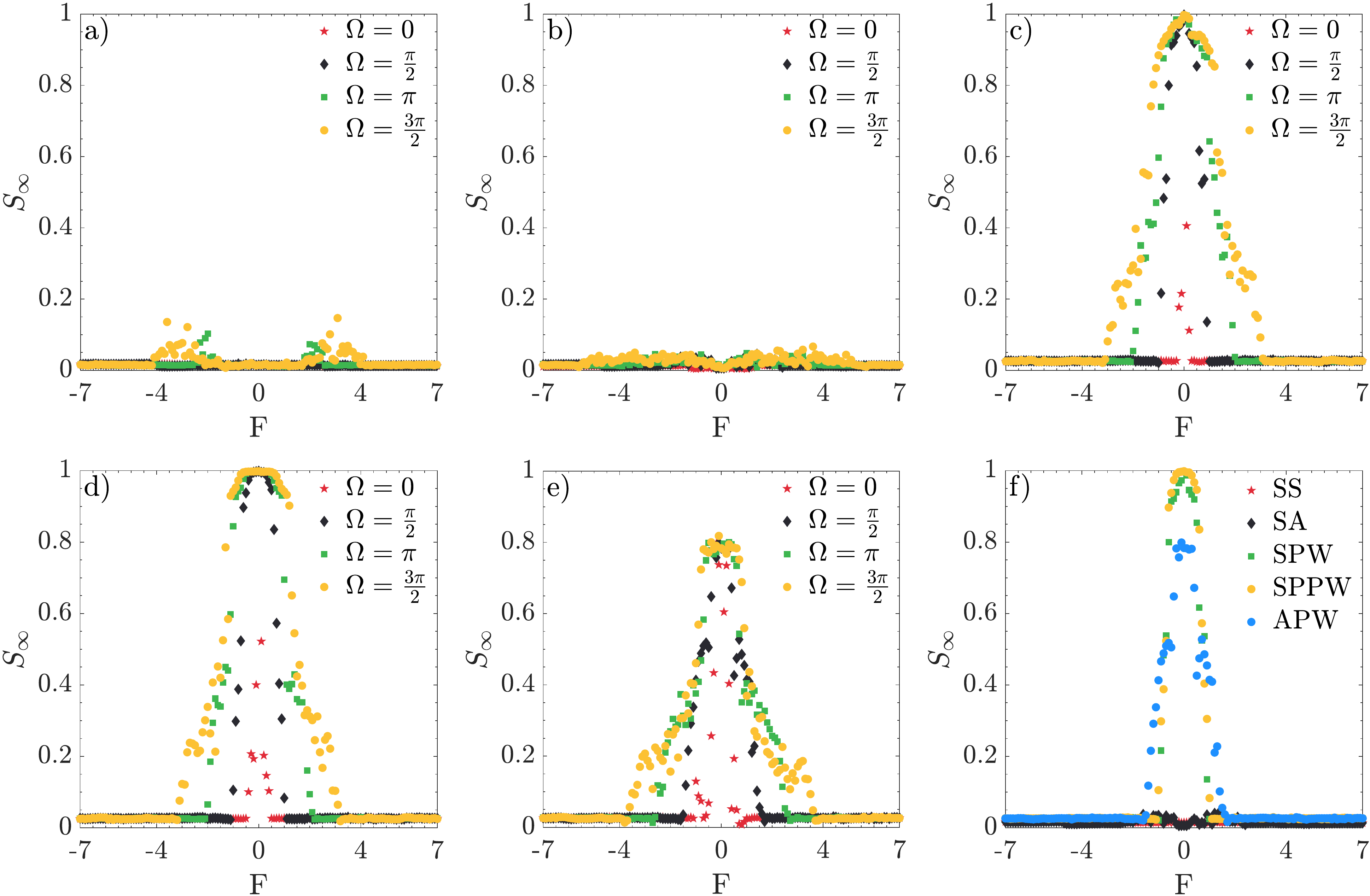}
\caption{\label{fig:correlation}Steady states of the correlation order parameter, represented as $S_{\infty}$, as a function of $F$ for a discrete set of frequencies $\Omega$ in each case. Graphs from \textit{a)} to \textit{e)} show phase transitions for the \textit{SS}, \textit{SA}, \textit{SPW}, \textit{SPpW} and \textit{APW} states respectively. Graph \textit{f)} shows a comparison between phase transitions generated by the five states for $\Omega= \pi/2.$}
\end{figure*}

\begin{figure*}
\includegraphics[scale= 0.385]{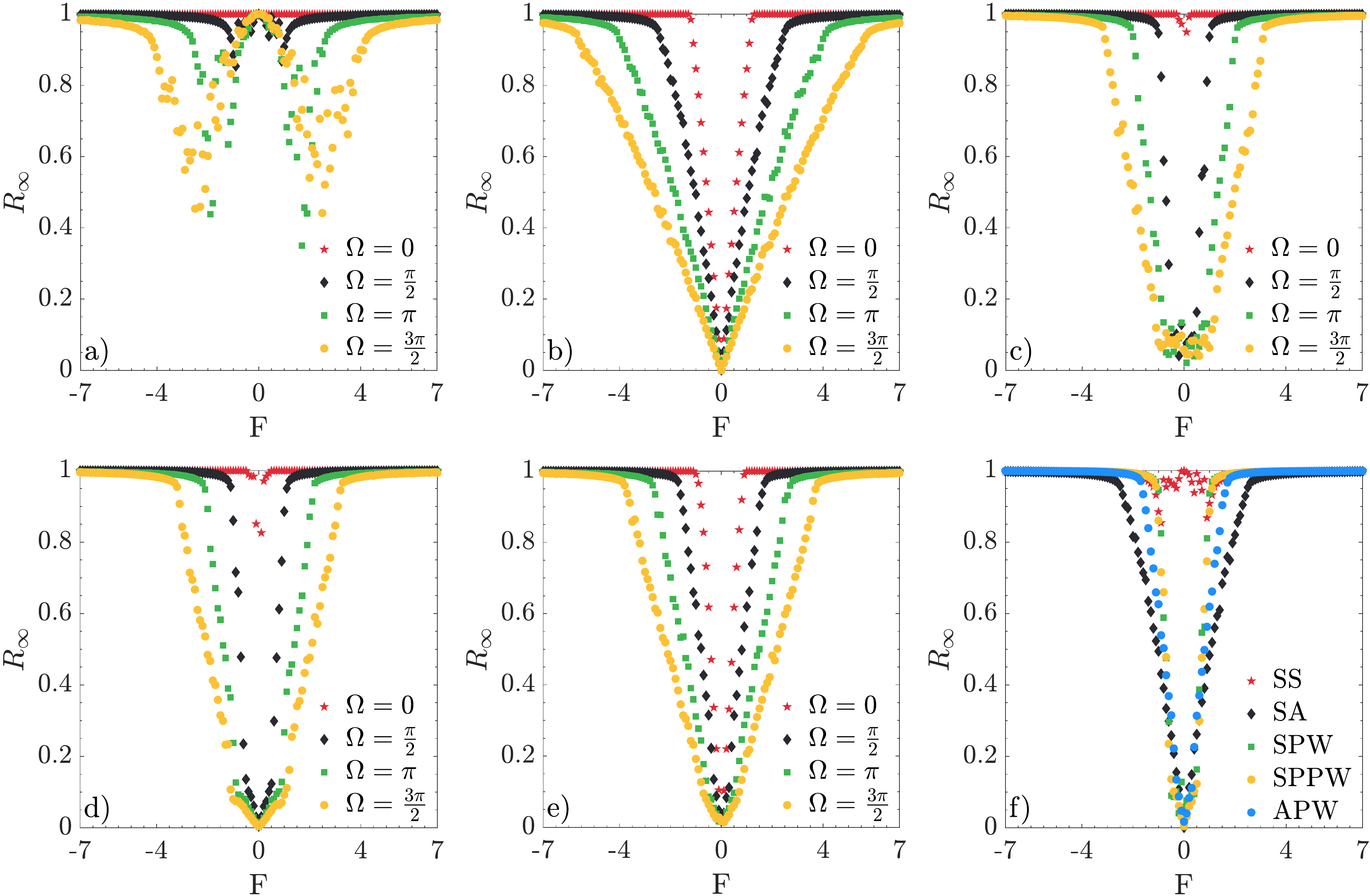}
\caption{\label{fig:phase_coherence}Steady states of the phase coherence order parameter, represented as $R_{\infty}$, as a function of $F$ for a discrete set of frequencies $\Omega$ in each case. Graphs from \textit{a)} to \textit{e)} show phase transitions for the \textit{SS}, \textit{SA}, \textit{SPW}, \textit{SPpW} and \textit{APW} states respectively. Graph \textit{f)} shows a comparison between phase transitions generated by the five states for $\Omega= \pi/2.$}
\end{figure*}

\subsection{\label{sec:Phase Transitions} Phase Transitions}

\begin{figure*}
\centering
\includegraphics[scale= 0.425]{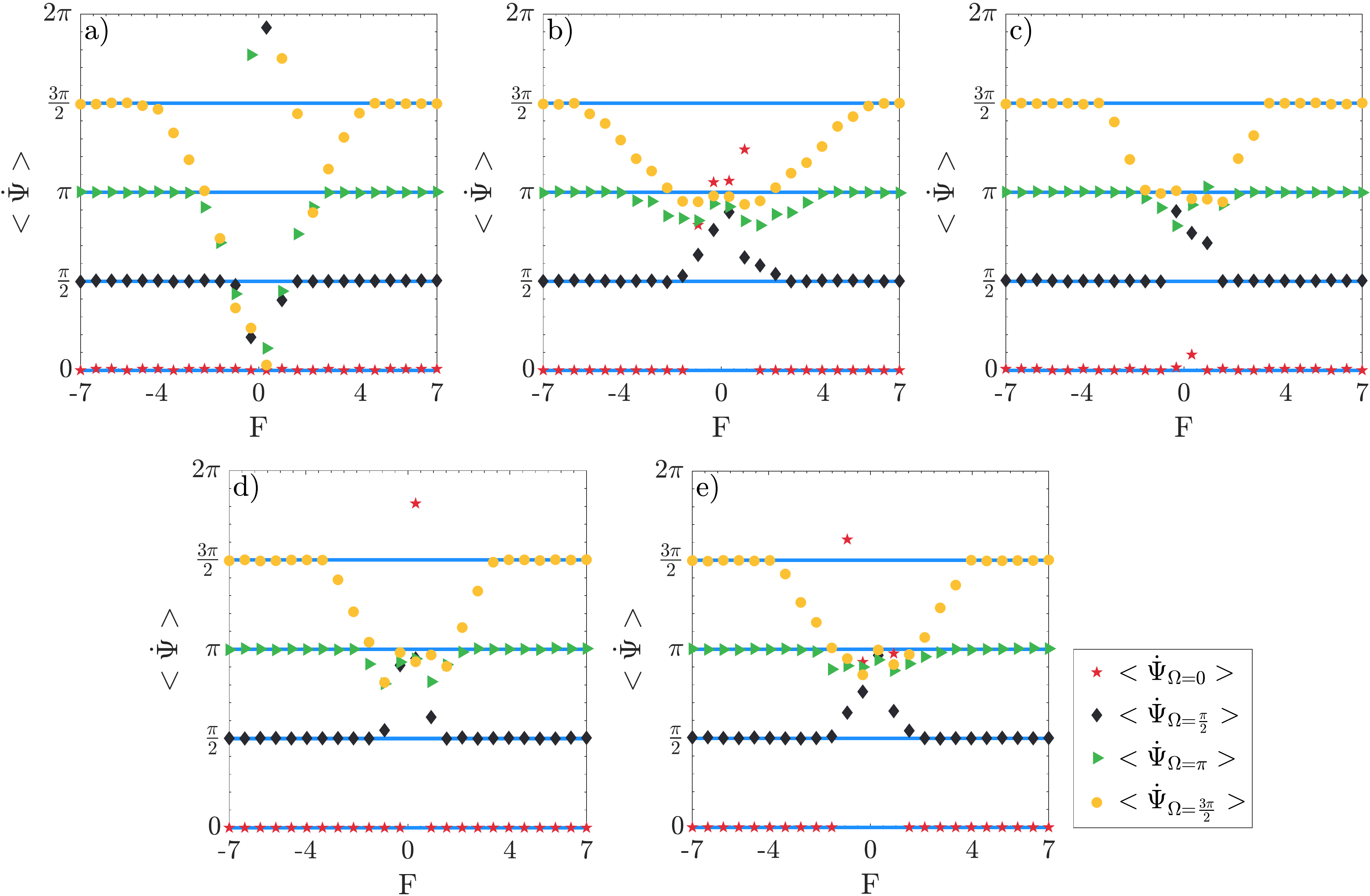}
\caption{\label{fig:locking} Transitions in the average of $\dot{\Psi}$ (represented as $<\dot{\Psi}>$) as a function of the amplitude $F$, for \textit{a) SS, b) SA, c) SPW, d) SPpW} and \textit{e) APW} cases. Stars, diamonds, triangles and circles represent $<\dot{\Psi}>$ for $\Omega= \{0, \pi/2, \pi, 3\pi/2\}$ respectively. Straight blue lines indicate stimulus frequencies $\Omega$ over $F$.}
\end{figure*}

Figs. \ref{fig:correlation} and \ref{fig:phase_coherence} show the behavior of the order parameters $R$ and $S$ in steady-state (defined as $R_{\infty}$ and $S_{\infty}$ respectively), with respect to a sweep of amplitudes $F$ from $-7$ to $7$ with steps of $0.1$, for frequencies $\Omega= \{0, \pi/2, \pi, 3\pi/2\}$.

Fig. \ref{fig:correlation} shows $S_{\infty}$, that measures the correlation between internal phases and spatial location. For the \textit{SPW}, \textit{SPpW} and \textit{APW} cases, $S_{\infty}$ changes sharply from their non-forced values ($S_{\infty}^{F= 0}$) when $F$ is turned on, for all frequency values $\Omega$. In contrast, for the two remaining cases \textit{SS} and \textit{SA}, $S_{\infty}$ displays only minimal variations for all the range of $F$. However, in all cases, after a critical value of the force, that depends on $\Omega$, $S_{\infty}$ goes to $0$ in what appears to be second order phase transition. This means that, independent of the state parameters, when the amplitude $F$ overcomes a threshold, phases and angular positions of the swarmalators become completely uncorrelated. 

Sharp transitions are also observed in the order parameter $R_\infty$, as shown in Fig. \ref{fig:phase_coherence}. Variations in $R_{\infty}$ as a function of $F$ are notorious in all the cases, as might be expected from the way the stimulus acts in the phase dynamics. If the correlation between free swarmalators is low (panels (b) to (e)), the stimulus forces them to synchronize with its own frequency, increasing  $R_{\infty}$. If, on the other hand, the free swarmalators are spontaneously synchronized (panel (a)), the stimulus initially forces them out of sync and then drives them back into sync with frequency $\Omega$. 

The behavior of $R_{\infty}$ indicates that, independent on the model parameters $(J,K)$, the introduction of a small external stimulus drives the system to a state of partial synchronization (except in the SS state where the original non-forced steady-state is fully synchronized). Moreover, there is a threshold in $F$, that depends on $\Omega$ and on $(J,K)$, that ensures full synchronization of the swarmalators with the external drive. As the threshold values for the phase transitions in $S_\infty$ and $R_\infty$ appear to be the same, we call it $F_{rc}$.

\subsection{\label{sec:Phase Locking} Phase Locking}

\begin{figure*}
\includegraphics[scale= 0.45]{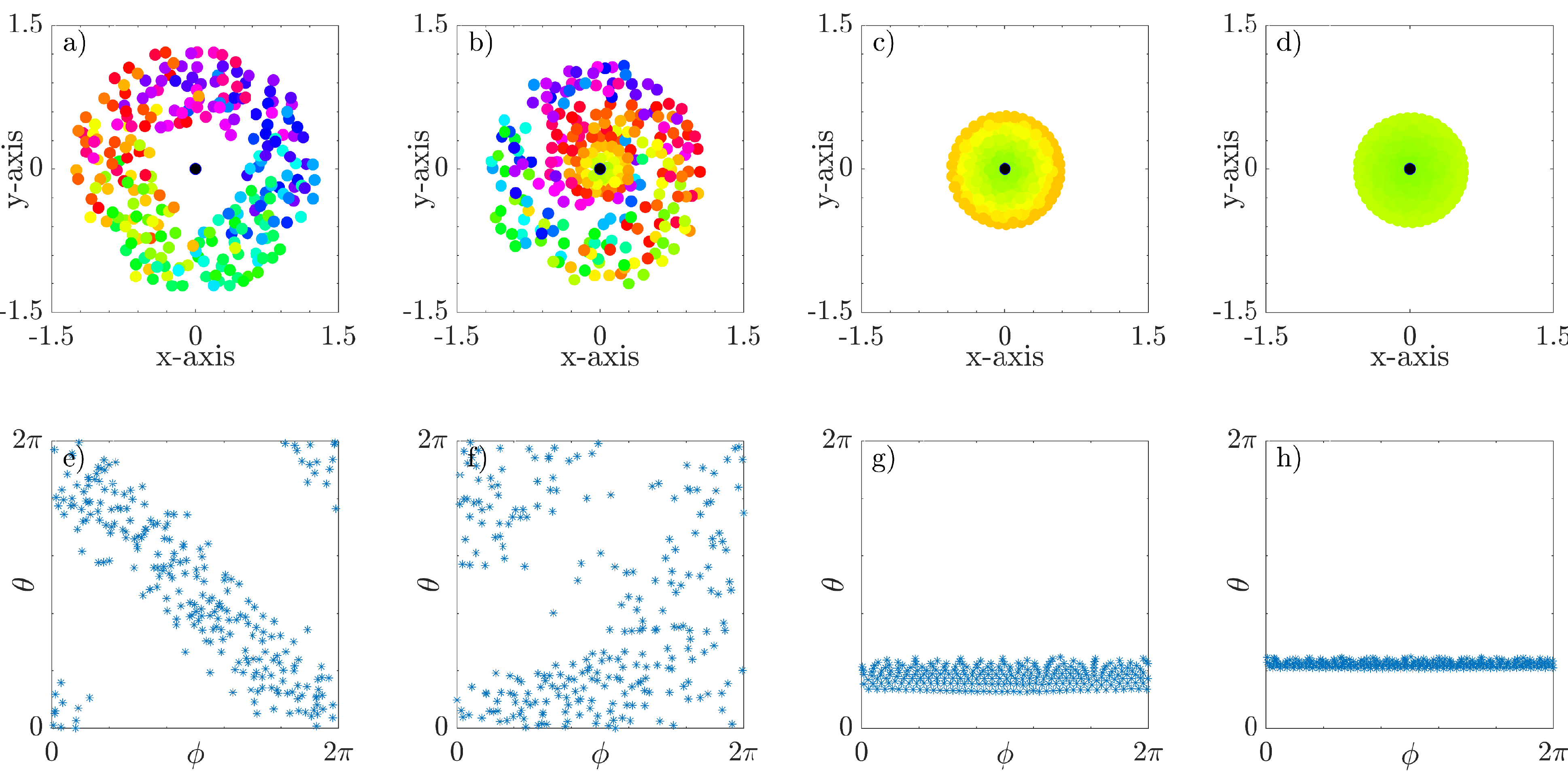}
\caption{\label{fig:particles} All the graphs in this panel are snapshots of the dynamic behavior of the system in steady state. Graphs in the top row show the spatial behavior of the system when order parameters $R$ and $S$ are in steady state, where the color of each particle represents its phase ($HSV$ representation with $H=0$, $S=0$, and $V=\theta_{i}$). Black dots in the origin represent the stimulus position. Simulations were performed for $\Omega= 3\pi/2$ and \textit{a)}$F= 0$, \textit{b)}$F= 1$, \textit{c)}$F= 1.7$ and \textit{d)}$F= 7$. Graphs in the bottom row (\textit{e)}, \textit{f)}, \textit{g)} and \textit{h)})show  scatter plots with respect to their pairs in the top row.}
\end{figure*}

The only non-forced state that shows synchronization in steady-state is the SS\cite{o2017oscillators}. Fig. \ref{fig:phase_coherence} shows that swarmalators also synchronize in steady-state when they are affected by amplitudes greater than $F_{rc}$. Thus, since the external stimulus is periodic we might expect that, just as in the forced Kuramoto model, the swarmalators synchronize their phases, while they also lock with the external stimulus frequency \cite{childs2008stability}.

Fig. \ref{fig:locking} shows how swarmalators' instantaneous phases variation evolve for different amplitudes $F$ and frequencies $\Omega$. We find that, independent on model parameters $(J, K)$, after the critical amplitude $F_{rc}$, the system synchronizes and locks with the external stimulus frequencies in all cases. In this regime the system also looses correlation between phases and angular positions. Thus, as shown in Fig. \ref{fig:particles}, we expect, in steady state, a radial uniform distribution of the particles in space. 



\section{Convergence to Fully Synchronous State}

Assuming that $F > F_{rc}$, the long term synchronization of the phases imply that $\theta_{j} \cong \theta_{i}$, leading to a simplification of Eq. (\ref{eq:dTheta}):
\begin{equation}\label{eq:simplified_phase}
    \dot{\theta}_{i}= \frac{F cos(\Omega t- \theta_{i})}{|\vec{x}_{0}- \vec{x}_{i}|}.
\end{equation}

In this approximation the phase of each oscillator is given by
\begin{equation}\label{eq:sync_rate}
    \theta_{i}= \Omega t + \alpha_i,
\end{equation}
where $F \cos{\alpha_i} = \Omega |\vec{x}_{0}- \vec{x}_{i}|$. This shows that, when the particles are fully synchronized with the external drive, their phases depend only on their distance to the source and have radial symmetry. Therefore, particles at the same radial distance get phase locked (see Fig. \ref{fig:transition}) and increase their velocity alignment. In the case of a ring-like spatial distribution of particles we can approximate $|\vec{x}_{0}- \vec{x}_{i}|$ by the ring radius and all particles become approximately phase-locked. In this case Eq. (\ref{eq:dX}) becomes
\begin{equation}\label{eq:simplified_dX}
    \dot{\vec{x}}_{i}= \frac{1}{N} \sum_{j\neq i}^{N} \bigg[\frac{\vec{x}_{j} - \vec{x}_{i}}{|{\vec{x}_{j} - \vec{x}_{i}}|} (1+ J)- \frac{\vec{x}_{j} - \vec{x}_{i}}{|{\vec{x}_{j} - \vec{x}_{i}}|^2} \bigg] 
\end{equation}
which represents a specific case of the Fetecau model of aggregation\cite{fetecau2011swarm}, where the trend of the particles is to be distributed uniformly inside of a circumference, just as in the case of static synchronization in the non-forced system.


\section{Discussion}

Swarmalators were originally defined as hypothetical entities that move in space and have an internal oscillatory degree of freedom, being able to swarm and sync \cite{o2017oscillators}. More recently, variations of this model have been proposed to describe cognitive systems\cite{monaco2019cognitive} and as an implementation of artificial systems\cite{hrabec2018velocity, o2019review, gniewek2019robots}. The nature of these applications reveals the importance of understanding their robustness, fidelity and response to external inputs.

In this work we studied the behavior of swarmalators under an external periodic driving. The stimulus is set up in a fixed position in space and affects particles' phases proportionally to their distance to source, similar to a flashing LED in a cloud of fireflies. We have shown that when the stimulus amplitude is increased from zero, each of the non-forced swarmalator's states experiences a transition from partial to full synchronization. Full coherence of swarmalators' phases occurs after a critical amplitude that is independent of the initial conditions of the particles and frequency of the stimulus. The full synchronization of the swarmalators is accompanied by a radial crystal-like distribution in space, drastically reducing the correlation between angular positions and phases. The second order transition shown in phase coherence as a function of the stimulus amplitude reminds of the emergent behavior towards synchronization present in the Kuramoto model.

The simulations show several qualitatively interesting features, such phase locking, spatial distribution and loss of correlation between internal and physical phases. Particularly important is the formation of the two-cluster state for $F=1$ and parameter values corresponding to \textit{SPW} and \textit{SPpW} states. This seems to be the only stable state that breaks the radial symmetry, although we have not explored the full space of parameters. We have also not obtained analytical expressions for the critical force $F_{rc}$ (as in the case of the forced Kuramoto model \cite{childs2008stability, moreira2019global, moreira2019modular}) or approximations that could explain the two-state pattern.

In our forced model we neglected the orientation of the particles and focused only on aggregation. Previous results revealed that orientation does not have strong influence in the long term states of the non-forced system \cite{o2017oscillators} . However, since the stimulus has an indirect effect over the spatial dynamics, the result might differ. We hypothesize that new interesting effects might be observed if the periodic external force were introduced to variations of the original model, such as the attractive phase coupling \cite{hong2018active}, or models where phase similarity affects spatial attraction and repulsion \cite{o2018ring}.

\begin{acknowledgments}
MAMA was partially supported by the Brazilian agencies CNPq (grant 302049/2015-0) and FAPESP (grant 2019/20271-5). We thank Flavia M. D. Marquitti and Luis F.P.P.F. Salles for the many comments and suggestions.
\end{acknowledgments}

\begin{appendix}
\section{Description of the movies}

All movies were obtained by numerically solving Eqs. (\ref{eq:dX}) and (\ref{eq:dTheta}) with $\Omega=3 \pi/2$ starting from random initial positions and phases. The force amplitude was varied from $F=0$ to $F=10$ at steps of $0.5$, except for movie 4 where $F=1$. The total integration time was 40,000 and force was increased by 0.5 at the end of each time interval of 2,000. The number of particles is $N=250$, except for movie 7 where $N=5$ .\\

\noindent Movie 1 - Transition from F = 0 to F = 10 for the SA state, $J=0.1$ and $K=-1$, with stimulus placed at the origin $(0, 0)$.\\

\noindent Movie 2 - Transition from $F = 0$ to $F = 10$ for the SPW state, $J=1$ and $K=0$ with stimulus placed at the origin $(0, 0)$.\\

\noindent Movie 3 - Transition from $F = 0$ to $F = 10$ for the SPpW state, $J=1$ and $K=-0.1$ with stimulus placed at the origin $(0, 0)$.\\

\noindent Movie 4 - Force amplitude fixed at $F=1$ for the SPW state with stimulus at (0, 0).\\

\noindent Movie 5 - Transition from $F = 0$ to $F = 10$ for the SPW state, $J=1$ and $K=-0$ with stimulus placed at $(0.5, 0.5)$.\\

\noindent Movie 6 - Transition from $F = 0$ to $F = 10$ for the APW state, $J=1$ and $K=-0.75$ with stimulus placed at $(0, 0)$.\\

\noindent Movie 7 - Transition from $F = 0$ to $F = 10$ for the APW state, $J=1$ and $K=-0.75$ with stimulus placed at $(0, 0)$ and $N=5$ particles.\\

\noindent Movie 8 - Transition from $F = 0$ to $F = 10$ for the SA state, $J=0.1$ and $K=-1$ and stimulus placed at $(0.5, 0.5)$.

\end{appendix}

\clearpage
\nocite{*}

\providecommand{\noopsort}[1]{}\providecommand{\singleletter}[1]{#1}%

\end{document}